\documentstyle[preprint,aps,tighten]{revtex}

\begin{document}                
\def\gsim{\mathrel{\rlap {\raise.6ex\hbox{$>$}}
{\lower.5ex\hbox{$\sim$}}}}
\def\lsim{\mathrel{\rlap {\raise.6ex\hbox{$<$}}
{\lower.5ex\hbox{$\sim$}}}}
\title{Search for Magnetic Monopoles Trapped in Matter}
\author{Hunmoo Jeon and Michael J.~Longo}
\address{University of Michigan, Department of Physics, Ann Arbor, MI 48109}
\address{(to be published in Physical Review Letters)}
\date{June, 1995}
\maketitle
\begin{abstract}                
There have been many searches for magnetic monopoles in flight, but few for
monopoles in matter.  We have searched for magnetic monopoles in meteorites,
schists, ferromanganese nodules, iron ores and other materials.  The detector
was a superconducting induction coil connected to a SQUID (Superconducting
Quantum Interference Device) with a room temperature bore 15 cm in diameter. We
tested a total of more than 331 kg of material including 112 kg of meteorites.
We found no monopole and conclude the overall monopole/nucleon ratio in the
samples is $<1.2 \times 10^{-29}$ with a 90\% confidence level.
\end{abstract}
%

\section{BACKGROUND}

In 1931
Dirac\cite{PAM}
suggested that the existence of magnetic monopoles can explain the quantization
of electric charge.  Dirac showed that a monopole must carry a magnetic
charge which is an integer multiple of $g_D = 68.5$ e in CGS units.  After
many unsuccessful searches for monopoles, interest gradually waned until the
Grand Unified Theory (GUT) magnetic monopole was introduced by t'
Hooft\cite{hooft} and independently by Polyakov\cite{poly} in 1974.  They
found that there is a solution with magnetic monopoles in any unified gauge
theory which has an unbroken U(1) factor embedded in it.  Since the U(1)
factor is essential for describing electromagnetism, virtually all unified
theories predict the existence of magnetic monopoles.  The most notable
property of the GUT monopoles is their extreme mass.  In all GUTs the
monopole mass ($m_M$) is of the order $\sim 4\pi m_X/e^2$ where {\it X} is
the mass of the superheavy vector boson, which determines the unification
energy scale\cite{four}.  For the SU(5) GUT, $m_X$ is $\sim 10^{14}$ GeV
which implies $m_M \approx 10^{16}$ GeV ($\sim0.01~\mu$g).  In SO(10),
monopoles with masses $\sim 10^4$ GeV could exist.  On the other hand
magnetic monopoles from GUTs with still larger groups, such as Kaluza-Klein
theory or supersymmetry generally have higher mass, $m_M
\gsim 10^{19}$ GeV.
 The magnetic charge of the GUT monopoles is predicted to be either 1 or 2
Dirac charges ($g_D$).

Because of the extreme mass of the GUT monopoles, a total rethinking of
search strategies was necessary.  It is generally agreed that GUT monopoles
can only have been created in the early ($\sim 10^{-35}$ sec) hot Big Bang
universe, possibly abundantly\cite{four}.  Thus the only GUT monopoles we can
possibly find are primordial.

Most of the experiments searching for them have been cosmic ray (or
in-flight) experiments\cite{five}.  An alternate, possibly more efficient way
is to search for magnetic monopoles trapped in matter.  In the Big-Bang
theory the GUT monopoles would be produced when the universe was $\sim
10^{-35}$ seconds old.  Hydrogen and helium synthesis began when the universe
was $\sim 1$ second old and ended when $\sim 10^2$ seconds old.  Stars and
galaxies are believed to form much later than that, when the universe is
$\gsim 10^{12}$ seconds old.  Heavy nuclei such as Fe and Ni are produced in
stars by nuclear fusion.  Magnetic monopoles, which might have been accreted
in the stars, should mostly have survived the evolution process because of
the exact conservation of magnetic charge.  Annihilations are inhibited by
even a small magnetic field and would not be expected to decrease the
monopole density by a large factor.  Eventually the monopoles may be recycled
in supernova, along with the heavy nuclei, to form new stars and planets.
Thus most of the magnetic monopoles may remain trapped in stars and other
astronomical bodies, probably near the core.  Presently the only macroscopic
samples we have of material from inside such bodies are meteorites and
meteoritic dust.

Monopoles can be bound to bulk ferromagnetic material by an image force.  For
iron (magnetite) the binding force near the boundary is $\sim 100$ eV/nm (60
eV/nm)\cite{six}, compared to the gravitational force $\sim 1$ eV/nm on a
monopole with mass $10^{17}$ GeV/c$^2$ near the surface of the Earth.  Thus a
monopole with mass $\lsim 10^{19}$ GeV/c$^2$ would be bound.  However if the
sample containing the monopole is subjected to an acceleration, the monopole
could be dislodged.  Meteorites entering the Earth's atmosphere undergo
accelerations of 100--1000 g.\cite{seven}  The interiors of meteorites remain
cold during their passage through the atmosphere.\cite{eight}  Any monopoles
they contain are likely to remain trapped in ferromagnetic grains in the
interior.  Thus the effective upper limit on the
monopole masses that meteorites might contain is $\sim 10^{16}-10^{17}$
GeV/c$^2$.  Monopoles also lose $\sim 0.02$ eV/nm in remagnetizing hard
magnetic material such as magnetite or meteoritic iron, so that monopoles
with mass $\lsim 10^{14}$ GeV/c$^2$ would only be extracted
by large sustained accelerations.

In general, primordial massive monopoles are unlikely to remain in material
near the Earth's surface, because most of the Earth's material was at one
time at a temperature well above the Curie point.  This means that any
primordial massive monopoles are likely to have fallen to the center.  If
there are magnetic monopoles with masses $\gsim 10^{16}$ GeV/c$^2$ in rocks
near the Earth's surface, they would
have to be captured after the rocks formed and cooled down below the Curie
temperature.  However it is easy to show that the chance of a massive
monopole being stopped near the Earth's surface is extremely low because the
range of a monopole is typically much greater than the diameter of the
Earth.\cite{nine}

Most meteors are fragments of small planets or asteroids.\cite{eight}  This
means that the meteorites falling on the Earth are samples from cores of
small planets which might contain monopoles.  Meteor ages are comparable to
that of the Earth.  Most have never been re-heated above the Curie point and
have a high content of iron.  This will insure that monopoles will reman
trapped in them via the monopole-image force if they have not been subjected
to overly large accelerations as explained above.

In addition to meteorites themselves, sedimentary rocks and ocean floor
sediments contain non-negligible amounts of meteoritic dust; this consists of
very small grains of meteoritic material constantly falling down on the Earth.
\cite{ten}  Typically $\sim 20~\mu$g of meteoritic dust is found in 100 g
of ordinary dust.\cite{eleven}  Also ferromanganese nodules found on the deep
sea bed have especially large numbers of meteoritic spherules rich in Fe and
Ni.\cite{twelve}  Another possibility is that monopoles could pass through
part of the Sun, losing enough kinetic energy to be captured in solar orbits
and eventually be stopped in the Earth.  While it is not easy to estimate the
possible monopole concentration due to these mechanisms, the rocks in the
Earth's crust are also of some interest in trapped monopole searches
considering their availability.

More than 20 years ago, Alvarez, Eberhard and collaborators\cite{thirteen}
tested $\sim 20$ kg of lunar rocks and other materials including $\sim 2$ kg
of meteorites.  Another somewhat similar experiment was done by Ebisu and
Watanabe.\cite{fourteen}  They tested iron ore from one of the deep mines in
Japan.  The iron ore was heated above the Curie temperature\cite{fifthteen}
and they looked for monopoles which fell from the ore.

\section{THE MONOPOLE DETECTOR}

Physically the monopole detector is a cylindrically-shaped vacuum chamber as
shown in Fig.~1.  Its length is 1295 mm and its diameter is 457 mm.  The
outer wall was made of mild steel to reduce the Earth's magnetic field.  A
mu-metal shield was placed just inside the outer wall for the same purpose.
The center hole (warm bore) is 152.4 mm in diameter and served as the test
path.  To reduce thermal noise (so-called Johnson noise), this was made of
G-10, an electrically non-conductive glass fiber/epoxy laminate.  A
liquid helium dewar, made of stainless steel, was placed concentrically.
Welded stainless steel is an ideal material for constructing cryostats;
however it is electrically conductive and, thus, a source of thermal noise.
In this case the thermal noise was acceptable, since the dewar was at
liquid helium temperature ($\sim 4.2^\circ$ K) and samples with ambiguous
monopole content could easily be passed through repeatedly.

The passage of a magnetic monopole
(trapped in a sample) will cause a jump in the persistent current in the
superconducting coil.  This minute change of current in the coil is measured
reliably with a SQUID.  For a superconducting coil with {\em N} turns and
inductance {\em L,} the change in current is $\Delta i = 4\pi Ng/L =
2\Delta i_o,$
where $\Delta i_o$ is the current change for a change of one flux quantum of
superconductivity, $\Phi_o = \hbar c/2e = 2.07 \times 10^{-7}$ G-cm$^2$.
Here we assume that the magnetic charge of a monopole is equal to 1 minimal
Dirac charge ($g_D = 2\Phi_o$).  A magnetic dipole will cause no net change
of magnetic flux and thus no persistent current in the coil.

To reduce vibration the detector was suspended by 3 nylon lines attached to the
3 ends of a T-shaped aluminum bar.  The bar was bolted to the ceiling via a
vibration damper.  Natural frequencies of the system were $\sim 1$ Hz
vertically and $\sim 0.5$ Hz horizontally.

Most of the samples we tested had a fair amount of ferromagnetic material,
and thus were strong magnetic dipoles in general.  Thus, with some samples
the SQUID output signal was changing even when the samples moved to the
outermost positions.  Sometimes these changes were indistinguishable from
that expected from a magnetic monopole.  This was mainly due the permanent
dipole moment of the sample, but sometimes the dipole moment induced by the
magnetic field outside the detector was as big as or bigger than the permanent
dipole moment.  The dipole signal was reduced significantly by placing the
sample in a tight fitting mu-metal container when necessary.

The expected response of the detector to a magnetic monopole was simulated
with a ``pseudopole", which is a very long solenoid with a diameter of 9.5 mm.
 The pseudopole was also used to verify that the SQUID measured the flux
charge properly when it was subjected to a huge transient dipole field.

The signal from the SQUID controller was digitized by an analog-to-digital
converter.  Internal feedback was not used in the SQUID controller, so the
electronics did not need to follow the large signals from dipole fields.  The
low pass filter in the SQUID controller was set to 1 Hz, and we digitized the
signal at a rate of $\sim 10$ per second.  The digitized signal was recorded
for 4 complete (up/down) cycles for each sample.  One cycle took $\sim 40$
seconds, or $\sim 160$ seconds to collect one set of data.  When the detector
had settled down, about 4 hours after filling with liquid helium, the noise
level was roughly one-third of the smallest Dirac charge (=2$\Phi_0$), as
simulated by the pseudopole.

\section{DISCUSSION AND CONCLUSIONS}

We tested meteorites and other samples, amounting to more than 743 in number
and 331 kg in total mass (Table I).  This is about 10 times the total
material tested by Alvarez, Eberhard et al.\cite{thirteen}  We did not find a
magnetic monopole.\cite{sixteen}  With these results, we can
conclude that the average overall ratio of the monopoles to nucleons in
our sample is $n_M/n_N < 1.2 \times 10^{-29}.$

If $m_M \gsim 10^{17}$ GeV, this experiment may not necessarily be a very good
test of the monopole/nucleon ratio of the universe because the trapped
monopoles could have been dislodged by accelerations which occurred during
the meteorites entry into the atmosphere, their impact, or subsequent
handling.

We wish to acknowledge the efforts of James Stone in early phases of the
experiment.  H.R. Gustafson provided considerable help and advice.  We
especially wish to thank J. Budai, E. Essene, D. Peacor, and Y. Zhang of the
University of Michigan Department of Geological Sciences for loaning us
many samples.  We are grateful to the NASA Meteorite Working Group for the
loan of 100 Antarctic meteorite samples.  This work was supported by the
National Science Foundation and grants from the Michigan Memorial Phoenix
Project and the Office of the Vice President for Research.

\pagebreak


\begin{table}
\caption{Summary of Samples}
\begin{tabular}{l|c|c|c|c} 
Samples\tablenotemark[1] &Mass &Number &Age &Notes \\ \hline
Perry collection meteorites\tablenotemark[2]
&32.7 kg &65 pcs &$>10^9$ years &2 \\
Antarctic meteorites &81.0 kg &99 pcs &$>10^9$ years &2 \\
Ferromanganese nodules\tablenotemark[3]
&5.0 kg &31 pcs &$10^2 - 10^7$ years &3 \\
Iron ores-magnetite\tablenotemark[4] &20.4 kg &39 pcs &$\sim10^9$ years &4 \\
Iron ores-hematite & 5.0 kg &8 pcs &$\sim 10^9$ years &4 \\
Blueschists\tablenotemark[5] &66.2 kg &77 pcs & $\sim 10^9$ years &5 \\
Sedimentary rock cores\tablenotemark[6] &112.5 kg
&$\approx$ 400 pcs &$\sim 10^9$ years &6 \\
Kimberlites\tablenotemark[7] &3.7 kg & 4 pcs & &7 \\
Chromates &4.9 kg &10 pcs & & \\
Garnets\tablenotemark[8] &$<$1 gm &$>$ 10 pcs &$\sim 10^9$ years &8\\ \hline
Total &331.4 kg &$>$743 pcs & & \\
\end{tabular}
\tablenotetext[1] {All the samples except for the Antarctic meteorites were
borrowed from the University of Michigan Department of Geological Sciences.
The Antarctic meteorites were borrowed from the NASA Meteorite Working Group.}
\tablenotetext[2] {Perry collection meteorites were collected from all over the
world.  Antarctic meteorites were collected from Antarctica.  Most of the
meteorites we tested are chondrites.}
\tablenotetext[3] {Ferromanganese nodules were collected from South Pacific
area (Clifferton-Clarion fracture zone).  Depth of the area is 4.5--5.5 km.}
\tablenotetext[4] {Magnetites were from mines all over the USA.  They are
ferrimagnetic.  Hematites were mostly from mines in Michigan state and are
anti-ferromagnetic.}
\tablenotetext[5] {Blueschists are scaly sedimentary rocks which contain modest
amounts of magnetite.  They formed at depths of 20--45 km and have never
experienced high temperature.}
\tablenotetext[6] {Sedimentary rock cores were drilled from 1.6 km depth,
Branch Co., Michigan.}
\tablenotetext[7] {Kimberlite are the base rocks in which diamonds are
imbedded.  They formed near the boundary between the earth's crust and mantle.}
\tablenotetext[8] {Garnets are also found in the blueschists.  The garnets are
formed at even higher pressure (thus depths) then the blueschists.}
\end{table}

\end{document}